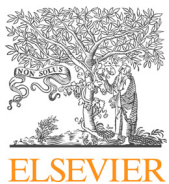

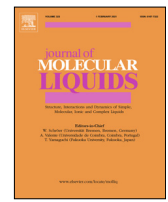

# Non-additive electronic polarizabilities of ionic liquids: Charge delocalization effects

Carlos Damián Rodríguez-Fernández [a,*], Elena López Lago [a], Christian Schröder [b], Luis M. Varela [a,*]

[a] NaFoMat Group, Departamento de Física Aplicada and Departamento de Física de Partículas, Universidade de Santiago de Compostela, E-15782 Santiago de Compostela, Spain
[b] University of Vienna, Department of Computational Biological Chemistry, Währingerstr. 17, A-1090 Vienna, Austria

ABSTRACT

Electronic charge delocalization on the molecular backbones of ionic liquid-forming ions substantially impacts their molecular polarizabilities. Density functional theory calculations of polarizabilities and volumes of many cations and anions are reported and applied to yield refractive indices of 1216 ionic liquids. A novel expression for the precise estimation of the molecular volumes of the ionic liquids from simulation data is also introduced, adding quadratic corrections to the usual sum of atomic volumes. Our significant findings include i) that the usual assumption of uniform, additive atomic polarizabilities is challenged when highly mobile electrons in conjugated systems are present, and ii) that cations with conjugated large carbon chains can be used together with anions for the design of ionic liquids with very high refractive indices. A novel relation for the polarizability volume is reported together with a refractive index map made up of the studied ionic liquids.

© 2021 The Author(s). Published by Elsevier B.V. This is an open access article under the CC BY license (http://creativecommons.org/licenses/by/4.0/).

## 1. Introduction

Ionic Liquids (ILs) are a set of materials composed solely of ions that are liquid at temperatures below 100 °C. Their exclusive ionic character and easy tunability provides them with a large set of properties that are useful in many different fields [1–12]. In particular, there is a growing interest in the optical properties of ILs as they have potential applications in photonics. In this regard, there are ILs showing liquid crystal behavior [13–15], luminescence [16–21], thermochromism [22,23], high refractive index [24] or nonlinear absorption and refraction [25,26]. Taking advantage of these properties, some authors went a step further by using ILs for specific photonic purposes. For instance, Tanabe et al. [27] designed a full-color luminescent tripodal ionic liquid crystal while Calixto et al. [28] produced an IL-based optofluidic compound lens. Moreover, Guo et al. exploited the thermo-optical effect of an-IL to all-optical control the attenuation effect on an optical fiber [29]. Yang et al. introduced them into a light-emitting electrochemical cell to increase its performance [30].

Notwithstanding, to fully implement ILs in the field of photonics, more profound knowledge of their optical properties is needed. Up to date, the most commonly measured optical properties of ILs are luminescence [17,19–21,27,31–36], refractive index at the sodium D line [37–49], and, less usually, refractive index dispersion [12,50–54] or non-linear optical response [25,26,55–60].

Nevertheless, all these studies only cover a reduced part of the vast number of possible combinations of cations and anions that yield photonically active ILs. In this context, several authors implemented different simulations and statistical procedures to predict optical properties of ILs, especially refractive index via their electronic polarizability [24,52,61–64]. These computational efforts are very diverse, and are based in methods ranging from ab initio density functional theory (DFT) calculations [24,65] to neural networks [64] or quantitative structure-property relationship models with predictive capability [61–63]. Most of these works consider electronic polarizability as a magnitude of linear character, which means that it is assumed that the electronic polarizability of each species can be calculated by the addition of uniform atomic components. These approaches are reasonable when dealing with systems where most electronic contributions come from $\sigma$ orbitals with a highly localized charge, and they can predict electronic polarizability with remarkable accuracy for particular sets of ILs. Nevertheless, it must be emphasized that electronic polarizability is not always a linear function of the number of each kind of atoms in a molecule. Indeed, it is well known that the existence of conjugated systems in a molecule, induced by the coupling of $\pi$ orbitals between adjacent atoms, produces a strong charge delocalization that leads to a non-linear behavior of the electronic polarizability.

---

* Corresponding authors.
*E-mail addresses:* damian.rodriguez@usc.es (C.D. Rodríguez-Fernández), luismiguel.varela@usc.es (L.M. Varela).

These substantial deviations of electronic polarizability from linearity have been studied both computationally [66–69] and experimentally [70–73] for different families of molecules presenting conjugated systems in their structures. However, despite the possible relevance of charge delocalization effects on the electronic polarizability of ILs was recently suggested [74], it has been scarcely studied up to the moment.

The objective of the current contribution is to provide new insights into the effect that charge delocalization plays in the electronic polarizability of ILs and how it impacts their refractive index. For this reason, the extent of charge delocalization induced by $\pi$ orbitals in a vast number of ions is studied, and significant trends are depicted. Specifically, the polarizability and volume of 92 different ions were simulated by DFT methods. These ions can be divided into a set of anions and three different groups of cations according to their structure (see Supplementary Material for consulting the molecular structures).

The first group of cations includes seven different cationic heterocyclic families with variable alkyl chain lengths. This group contains highly aromatic cations such as 1-alkylquinolinium [$C_k$quin]$^+$ or 1-alkylpyridinium [$C_k$py]$^+$ as well as not aromatic cations such as 1-alkyl-1-methylpyrrolidinium [$C_k$mpyrr]$^+$ or 1-alkyl-1-methylpiperidinium [$C_k$mpip]$^+$. The presence or absence of aromaticity in these different IL families is expected to provide valuable information on the influence of charge delocalization on their electronic polarizability and refractive index.

The second cationic group includes diverse singly and doubly substituted imidazolium species containing different number and positions of $\pi$ bonds, such as 1-allyl-3-methylimidazolium [allylmim]$^+$, 1,3-diallylimidazolium [1,3-diallylim]$^+$, 1-benzyl-3-methylimidazolium [benzylmim]$^+$, 1,3-dibenzylimidazolium [1,3-dibenzylim]$^+$, 1-crotyl-3-methylimidazolium [crotylmim]$^+$, 1-ethoxy-3-methylimidazolium [eomim]$^+$, 1,3-diethoxyimidazolium [1,3-dieoim]$^+$, 1-ethylnitrile-3-methylimidazolium [enmim]$^+$, 1,3-diethylnitrileimidazolium [1,3-dienim]$^+$, 1-methyl-3-vinylimidazolium [vinylmim]$^+$, 1,2-divinylimidazolium [1,2-divinylim]$^+$ and 1,3-divinylimidazolium [1,3-divinylim]$^+$.

The last group comprises imidazolium-based cations with different kinds of functionalized side chains, e.g. 1-meth(oxyethyl)$_k$-3-methylimidazolium [m(eo)$_k$mim]$^+$, 1-methyl-3-perfluoroalkyli midazolium [$F_k$mim]$^+$ and 1-polyenyl-3-methylimidazolium [uC$_k$-mim]$^+$. The alkyl chain serves as a reference, while perfluorinated and oxygenated chains are interesting since their basic units contain elements that go beyond carbon and hydrogen. Lastly, the polyenyl side chain presents a totally delocalized structure produced by the conjugation of the $\pi$ bonds throughout the chain. It will provide the clearest picture of how charge delocalization influences polarizability.

The polarizability and volume of the different ions were treated to accurately estimate the refractive index of all their possible combinations. Thus, we present here estimations of the refractive index of 1216 ionic pairs, some of them never synthesized. The model developed to move from the individual polarizability and volume of the ions to that of their combination provides extra insights into how the ion combination affects the resulting refractive index, pointing to new strategies to reach ILs with refractive indices higher than 2.0. This model includes a quadratic volume correction that highly improves the theoretical IL volume calculated by the raw addition of the quantum mechanical volumes of the ions. This correction can improve the prediction of other volume-based magnitudes, such as mass density, using data from ab initio calculations.

## 2. Theoretical and computational details

### 2.1. Selection of the DFT functional and basis set

Ab initio DFT calculations were carried out using the Gaussian 16 rev. C.01 program [75]. All calculations were performed for isolated ions whose geometry was properly optimized and tested employing a standard vibrational analysis. On the equilibrium structures, electronic polarizability in the range from 400 nm to 1500 nm was calculated using the Coupled-Perturbed Hartree-Fock (CPHF) method [76]. Molecular volume was defined as the volume inside a contour of 0.001 electrons/Bohr$^3$ according to [77] and calculated utilizing the Monte-Carlo procedure available in Gaussian. Since the algorithm can provide significantly different results after each execution [78], the volume calculation was repeated 100 times for each ion to obtain good statistics. The level of theory employed, B3LYP/6-311++G(d,p), was chosen after testing the performance of different potentials and basis sets combinations in the prediction of electronic polarizability and refractive index dispersion of a set of reference ILs. The tested potentials were B3LYP, CAM-B3LYP, M062X and PBE0 in combination with the 6-311++G(d,p) and aug-cc-pVDZ basis sets, and the studied ILs were 1-alkyl-3-methylimidazolium tetrafluoroborate [$C_k$mim][BF$_4$] with $k$ = 2, 3, 4, 6 and 8, 1-alkyl-3-methylimidazolium bis(tri fluoromethane)sulfonylimide [$C_k$mim][NTf$_2$] with $k$ = 2, 3, 4 and 6, 1-ethyl-3-methylimidazolium trifluoromethanesulfonate [$C_2$mim][OTf] and 1-ethyl-3-methylimidazolium alkylsulfate [$C_2$mim][$C_k$-SO$_4$] with $k$ = 2 and 6. The experimental electronic polarizability of the test ILs was calculated using our previously reported data [50] by means of the Lorentz-Lorenz equation:

$$\frac{n^2 - 1}{n^2 + 2} = \frac{4\pi}{3} \frac{\alpha_{IL}}{V_{IL}}, \tag{1}$$

where $V_{IL}$ is the molecular volume of the IL and $\alpha_{IL}$ its electronic polarizability volume. Please, note that, for convenience, the electronic polarizability volume is used instead of the actual electronic polarizability, being both magnitudes related by a $4\pi\varepsilon_0$ factor with $\varepsilon_0 = 8.85 \times 10^{-12}$ C V$^{-1}$ m$^{-1}$, the vacuum electric permittivity. The simulated electronic polarizability of the respective ILs was calculated by adding the electronic polarizability contributions of their composing ions $\alpha_{IL} = \sum_i \alpha_i$. This form of calculating the total elec-tronic polarizability is quite convenient since it provides flexibility and speed to our calculations. Nevertheless, it has to be managed carefully since additivity on electronic polarizability is not always ensured. In this regard, later in the text, there is an in-depth discussion about the actual extent of the linear behavior of electronic polarizability on ILs. In this specific case, since no charge transfer in the optical regime of energies is expected, it seems reasonable to assume it. This assumption is supported by both available bibliography [62,63,65] and our experimental results, which are shown in Table 1. The relative error in this table, $\Delta \alpha = (\alpha_{IL}^{sim} - \alpha_{IL}^{exp})/\alpha_{IL}^{exp}$, is defined with respect to experimental measurements and averaged over the spectrum and members of each family of tested ILs. It is below 5% for most of the combinations, and the most accurate combination is the B3LYP/6-31++G(d,p) with relative errors below 2%.

**Table 1**
Relative error of simulated electronic polarizability with respect to experimental data [50]. The tabulated values correspond to the average of the relative error over spectra and members of each family of ILs.

| | | Electronic polarizability relative error, $\Delta\alpha$ | | | |
|---|---|---|---|---|---|
| Functional | Basis set | [$C_k$mim][BF$_4$] | [$C_k$mim][NTf$_2$] | [$C_2$mim][OTf] | [$C_2$mim][$C_k$SO$_4$] |
| B3LYP | 6-31++G(d,p) | 1.2% | 1.7% | 0.5% | 1.6% |
| | aug-cc-pVDZ | 6.1% | 9.0% | 2.4% | 8.4% |
| CAM-B3LYP | 6-31++G(d,p) | 3.1% | 2.2% | 3.5% | 1.8% |
| | aug-cc-pVDZ | 3.4% | 4.9% | 3.8% | 4.8% |
| M062X | 6-31++G(d,p) | 4.1% | 3.7% | 5.1% | 3.2% |
| | aug-cc-pVDZ | 2.6% | 3.2% | 2.1% | 3.8% |
| PBE0 | 6-31++G(d,p) | 2.1% | 1.1% | 2.8% | 7.3% |
| | aug-cc-pVDZ | 4.3% | 5.9% | 4.4% | 5.8% |

### 2.2. The effect of charge delocalization

It is possible to obtain a closer insight into the microscopic origins of electronic polarizability through quantum mechanical calculations. A simple model of the electrons in a molecule under the effect of an electric field is a free electron gas composed of 2 N electrons confined in a one-dimensional box of length *L* [66,79]. Its global polarizability can be calculated as the sum of the contributions from each excited state $\xi$, which can be analytically calculated as the second partial derivative of the energy of that state $E_\xi$ with respect to the perturbing electric field *F* [66]:

$$\alpha_i = -2 \sum_{\xi=1}^{N} \frac{\partial^2 E_\xi}{\partial F^2} = \frac{4L^4}{a_0} \sum_{\xi=1}^{N} \left( \frac{-2}{3\pi^2 \xi^2} + \frac{10}{\pi^4 \xi^4} \right) \tag{2}$$

where $a_0 = \hbar^2/me^2$ is the atomic Bohr radius. From Eq. (2), it is observed that the electronic polarizability presents a fourth-power dependence on the length of the box, that is, the physical extension where the electrons are allowed to move following the electric field. The employment of more realistic approaches to this problem yields different analytical forms for the electronic polarizabilities; however, a power dependence on *L* is always retained [79,80]. As the effective *L* where conjugated electrons are allowed to move is larger than that of localized ones, the increase of the size of the delocalization region in a molecule is expected to produce a non-linear increase of the associated electronic polarizability. We have investigated this effect by studying the polarizability of imidazolium cations [uC$_k$mim]$^+$ with conjugated side chains of variable length (*k*).

For the computation of the atomic polarizabilities of this conjugated family, $\alpha_\beta$, we employed the methods described in [81,82]. The average atomic polarizability of an atom $\beta$ is obtained as the numerical derivative of the atomic dipole moment $\mu_\beta$ with respect to an applied electric field *F*,

$$\alpha_\beta = \frac{1}{3} \left( \frac{\partial \mu_{\beta x}}{\partial F_x} \bigg|_{F_x=0} + \frac{\partial \mu_{\beta y}}{\partial F_y} \bigg|_{F_y=0} + \frac{\partial \mu_{\beta z}}{\partial F_z} \bigg|_{F_z=0} \right) \tag{3}$$

The atomic dipoles $\mu_\beta$ are the net dipole moment in a basin $\Omega_\beta$ arising from the anisotropy of the electron cloud around the nucleus and a charge transfer term,

$$\vec{\mu}_\beta = \vec{\mu}_\beta^p + \vec{\mu}_\beta^{CT} \tag{4}$$

$$\vec{\mu}_\beta^p = \int_{\Omega_\beta} \rho\left(\vec{r}\right) \cdot \left(\vec{r} - \vec{R}_\beta\right) d\vec{r} \tag{5}$$

$$\vec{\mu}_\beta^{CT} = \sum_\gamma q_{b(\beta\gamma)} \cdot \left(\vec{R}_\beta - \vec{R}_{b(\beta\gamma)}\right) \tag{6}$$

The first term $\vec{\mu}_\beta^p$ is called the polarization term and can be evaluated by the GDMA code of Misquitta and Stone [83,84]. It contains the electron density $\rho\left(\vec{r}\right)$ and is defined with respect to the coordinate $\vec{R}_\beta$ of the nucleus $\beta$. This contribution describes the local deformability of the electron cloud. The second contribution arises from the charge transfer (CT) between the bonds $b(\beta\gamma)$ between the nuclei $\beta$ and $\gamma$. The bond charge $q_{b(\beta\gamma)}$ can be defined in such a manner [82] that it does not depend on the origin of the coordinate system (even for charged molecules). It is located at the center between the nuclei $\vec{R}_{b(\beta\gamma)} = 1/2 \cdot \left(\vec{R}_\beta + \vec{R}_\gamma\right)$. The procedure to obtain these atomic dipole moments and their constituting parts is detailed in [24,81,82]. For the discussion in this work, the polarizability in Eq. (3) can also be decomposed into a polarization and a charge transfer term:

$$\alpha_\beta = \alpha_\beta^P + \alpha_\beta^{CT} \tag{7}$$

Intuitively, the charge transfer contribution should strongly depend on the nature of the bound system. In other words, we expect high $\alpha_\beta^{CT}$ as a function of the length of the conjugated system. Please note that the sum of the atomic polarizabilities $\alpha_\beta$ is still the molecular polarizability $\alpha_i$ applying this algorithm:

$$\alpha_i = \sum_\beta \alpha_\beta \tag{8}$$

However, this does not cast doubt on the polarizability's non-linearity as individual atomic polarizabilities $\alpha_\beta$ are not uniform for particular atom types such as sp$^2$ carbons. In other words, individual atomic polarizabilities $\alpha_\beta$ may still be a function of the size of the electron delocalization.

In order to evaluate the amount of charge delocalization within the studied ions, we employed the Multiple Center Bond Order (MCBO) index [85], which is a quantitative estimator of the degree of charge delocalization in molecules. The MCBO index was calculated in all the cases using the Multiwfn 3.8 program [86]. As we used it to compare molecular fragments of a different number of atoms, we employed here the normalized version of this parameter [87], and we used as final MCBO values the average resulting from considering the two opposite directions of the path including the atoms of interest. The absolute value of MCBO is a quantitative parameter that measures the extent of charge delocalization in a molecular structure. The higher it is, the more significant is the extent of charge delocalization. The sign of the MCBO is related to the orbital origin of the electrons participating in the delocalization [88].

### 2.3. Calculation of the molecular volume to evaluate the refractive index

Although simulating electronic polarizability and refractive index is related, the refractive index's calculation also requires estimating the IL molecular volume, see Eq. (1). However, the IL molecular volume, $V_{IL}$, is larger than that of the mere addition of the volumes of its constituent ions, $V_{IL} > \sum_i V_i$, a behaviour that was already reported in recent works [82]. For this reason, we decided to use a semi-empirical expression to estimate from the ion volumes, $V_i$, the ionic liquid volume, $V_{IL}$. Different combination rules were evaluated to find the expression for $V_{IL}$ which better reproduces the experimental refractive index of the test ILs. We concluded that a quadratic correction in terms of the sum of ionic volumes ($\sum_i V_i$) with two fitting parameters, $f_{scale}$ and $f_{int}$, is enough to obtain representative IL volumes:

$$V_{IL} = f_{scale} \cdot \sum_i V_i + f_{int} \cdot \left( \sum_i V_i \right)^2 \tag{9}$$

It is possible to obtain some physical insight into the semi-empirical *f*-factors in the above equation. $f_{scale}$ is a factor whose value is always > 1, suggesting a generalized underestimation on the calculation of the molecular volume of the ions. According to the literature [77,89], this underestimation is produced because the real volume that a molecule occupies in the condensed phase (the cavity volume) is up to 30 % larger than the molecular one. On the other hand, the second parameter $f_{int}$ summarizes various interactions and packing effects. It can be interpreted as an interaction or excess volume coefficient since it contains quadratic terms arising both from self and cross-interactions. Both factors could also be slightly dependent on the temperature mismatch between experimental data (obtained at room temperature) and calculations (at 0 K). In addition, one often neglects that ionic liquids' ion cages consist of a more or less equal number of cations and anions [90]. It is also possible that the number of cations around a central cation exceeds the number of anions, which is more probable when increasing the side chain length, *k*. For example, imidazoliums with $k > 6$ tend to form cationic micelles with large hydrophobic regions and anion depletion. In these aggregates, the Coulombic repulsion between like-charged ions increases the volume in the liquid phase. The larger the central cation volume, the larger the number of nearest neighbouring cations, which may explain the quadratic influence.

The fitting parameters in Eq. (9), $f_{scale}$ and $f_{int}$, are dependent on the level of theory used and, for this reason, they were evaluated for all the DFT functionals and basis sets considered. Table 2 shows the performance of our correction for different levels of theory, being $\Delta n = (n_{sim} - n_{exp})/n_{exp}$ the average relative error of our simulation results with respect to experimental ones [50]. The best predictions of refractive indices are those of the B3LYP/6-311++G(d,p) level of theory, as in the polarizability case, which provides uncertainties in the refractive index in the wavelength range of $U(n) \approx 1 \cdot 10^{-2}$ with $f_{scale} = 1.0189$ and $f_{int} = 3.635 \times 10^{-4}$ Å$^{-3}$. For this reason, it is the level of theory chosen for the performance of the calculations presented in this work. Except indicated otherwise, all given polarizabilities and refractive indices were calculated at a wavelength of $\lambda = 589$ nm.

As the value of the refractive index *n* of the ILs considered in this study is limited between 1.35 ([F$_5$mim][FAP]) and 1.69 ([uC$_5$-mim][SCN]), the Lorentz-Lorenz equation, Eq. (1), can be approximated with a simple Taylor series [91]:

$$\frac{4\pi}{3} \frac{\alpha_{IL}}{V_{IL}} = \frac{n^2-1}{n^2+2} \approx c_1 \cdot n + c_2 \qquad (10)$$

Thus, the overall polarizability density $\alpha_{IL}/V_{IL}$ is roughly a linear function of the refractive index and vice versa. In an ideal mixture, one could expect that the $\alpha_{IL}/V_{IL}$ ratio, and hence, the refractive index, were directly obtained as the usual linear combination of the individual molecular polarizability density of each *i* ion:

$$\left(\frac{\alpha_{IL}}{V_{IL}}\right)^{ideal} = \sum_i \text{MPD}_i \cdot \phi_i, \qquad (11)$$

where the molecular polarizability density, MPD$_i$, is defined as:

$$\text{MPD}_i = \frac{\alpha_i}{V_i^{\text{MPD}}}, \qquad (12)$$

and $\phi_i = V_i^{\text{MPD}}/\sum V_i^{\text{MPD}}$ is the volume fraction of the species *i* after correcting it by means of Eq. (9), $V_i^{\text{MPD}} = f_{scale} \cdot V_i + f_{int} \cdot V_i^2$. However, since $V_{IL}$ involves cross-terms between the volume of its composing ions, the previous linear combination is just an upper limit for the real $\alpha_{IL}/V_{IL}$ ratio and, thus, IL refractive index:

$$\left(\frac{\alpha_{IL}}{V_{IL}}\right)^{real} \leqslant \sum_i \text{MPD}_i \cdot \phi_i. \qquad (13)$$

## 3. Results and discussion

### 3.1. Charge delocalization and polarizability on the anion

Naturally, ILs consist of cations and anions. However, most common ion combinations involve anions that do not have large π systems, except tosylate. Some of the most common anions such as acetate, trifluoroacetate, methylsulfate, methylsulfonate, triflate, or bis(trifluoromethane)sulfonimide possess double bonds. Cyanide-based anions like thiocyanide, dicyanamide, and tetracyanoborate have triple bonds. Acetates, cyanides and methylsulfonyl-based anions have mesomeric structures indicating a delocalized charge region centered at the carbon in acetate and cyanide anions, and in the nitrogen in methylsulfonyl anions. Table 3 shows the MCBO and the MPD$_i$ of some of the considered anions. The MCBO parameter is a quantitative measurement of the degree of charge delocalization in a molecule, being the [SCN]$^-$ anion the one with the highest value of this parameter. In fact, its MCBO is even higher than that of tosylate, whose value is also lower than that of other anions showing resonance. Nevertheless, both anions present the highest polarizability per unit volume, which we define as molecular polarizability density MPD$_i$, see Eq. (12). Concerning the polarizabilities $\alpha_\beta$ on the atomic level, the situation is less clear. For example, the sp$^2$ carbons in tosylate show the same polarizability as that of the sp$^2$ carbon in acetate despite the extensive delocalized system. The only exception is the carbon bound to the sulfur in tosylate, which has a significantly higher polarizability of 1.75 Å$^3$. The immediate neighborhood to a highly polarizable sulfur atom also leads to high polarizability of the sp carbon in thiocyanate of $\alpha_\beta = 2.16$ Å$^3$, which is significantly higher than the polarizability of the carbons in the other cyanides. The presence of sulfur also drives the polarizability of the nitrogen in [NTf$_2$]$^-$. The overall ranking on the MPD$_i$ is depicted in Fig. 1. Anions containing many fluorine atoms have lower ratios and are located on the left end of the plot. In the case of [OTf]$^-$ and [NTf$_2$]$^-$, the low polarizability of the fluorine is compensated to some extent by the more polarizable sulfur, carbon and oxygen atoms. The only metal-containing anion considered in this work, the

**Table 2**
Relative error of simulated refractive index, $\Delta n$, with respect to experimental data [50]. The tabulated values correspond to the average of the relative error over spectra and members of each family of ILs.

| | | Refractive index relative error, $\Delta n$ | | | |
|---|---|---|---|---|---|
| Functional | basis set | [C$_6$mim][BF$_4$] | [C$_6$mim][NTf$_2$] | [C$_2$mim][OTf] | [C$_2$mim][C$_8$SO$_4$] |
| B3LYP | 6-31++G(d,p) | 0.2% | 0.7% | 0.5% | 1.1% |
| | aug-cc-pVDZ | 2.7% | 3.3% | 0.1% | 0.4% |
| CAM-B3LYP | 6-31++G(d,p) | 1.0% | 0.9% | 1.1% | 3.1% |
| | aug-cc-pVDZ | 1.1% | 0.8% | 4.6% | 0.5% |
| M062X | 6-31++G(d,p) | 0.5% | 0.6% | 0.4% | 1.4 % |
| | aug-cc-pVDZ | 1.3% | 1.7% | 2.6% | 1.1% |
| PBE0 | 6-31++G(d,p) | 0.9% | 0.8% | 1.2% | 2.3% |
| | aug-cc-pVDZ | 1.2% | 2.4 % | 2.3% | 1.2% |

**Table 3**

Normalized Multi-Center Bond Order (MCBO) and MPD$_i$ of the anions. The atomic polarizabilities $\alpha_\beta$ are taken from [91].

| Anion | \|MCBO\| | MPD$_i$ | $\alpha_\beta$ (N) [Å³] | $\alpha_\beta$ (C$_{sp}$) [Å³] | $\alpha_\beta$ (C$_{sp}$) [Å³] | $\alpha_\beta$ (C$_{sp}$) |
|---|---|---|---|---|---|---|
| [OTf]⁻ | 0.17 | 0.0609 | | | | 0.98 |
| [NTf$_2$]⁻ | 0.23 | 0.0641 | 1.99 | | | 1.03 |
| [B(CN)$_4$]⁻ | 0.26 | 0.0703 | 1.31 | 1.23 | | |
| [C$_1$SO$_4$]⁻ | 0.31 | 0.0660 | | | | 0.88 |
| [Tos]⁻ | 0.40 | 0.0856 | | | 1.29–1.52,1.75 | 1.10 |
| [C$_1$SO$_3$]⁻ | 0.47 | 0.0686 | | | | 1.12 |
| [N(CN)$_2$]⁻ | 0.49 | 0.0816 | 1.78 | 1.46 | | |
| [OAc]⁻ | 0.51 | 0.0796 | | | 1.34 | 1.13 |
| [SCN]⁻ | 0.73 | 0.0863 | 1.88 | 2.16 | | |

[AlCl$_4$]⁻ has an MPD$_i$ which is not exceptionally high or low, but similar to that of other considered anions. As aluminium is a closed-shell cation, its optical absorption is not expected to be intense; however, the introduction of other kind of metallic complexes with absorption bands closer to the visible range could produce anions with higher MDP$_i$. It is also interesting to note that the introduction of an extra oxygen atom in [C$_1$SO$_3$]⁻ to form [C$_1$SO$_4$]⁻, results in a decrease of the MPD$_i$. The anions with some degree of delocalization, *e.g.* acetate, dicyanamide, tosylate, and thiocyanate have top ratios and are located at the right end of the plot in Fig. 1.

### 3.2. Influence of the cationic core on the polarizability

We start with the analysis of the first group of cations, those containing several aromatic and aliphatic cores and aliphatic side chains. Their polarizability volume $\alpha_i$ versus their molecular volume $V_i^{MPD}$ is depicted in Fig. 2a. Prolonging the side chain increases both the molecular polarizability and the cation volume, resulting in a linear trend for all cationic species sharing more or less the same slope. These similar slopes indicate that electronic polarizability behaves linearly upon addition of the $\sigma$ electrons of the −CH$_2$− group. However, the corresponding intercepts in Fig. 2a differ. On the one hand, this may be due to the different nature of the cationic cores. Although all these cations contain mainly carbons and nitrogens with comparable atomic polarizabilities [62,63], their hybridization varies between sp² and sp³. Additionally, 1-alkyl-3-methylthiazolium [C$_k$mthia]⁺ contains sulfur and 1-alkyl-1-methylmorpholinium [C$_k$mmor]⁺ has one oxygen.

However, the largest intercept is found for 1-alkylquinolinium [C$_k$quin]⁺ which is the largest cation under investigation and also has the most extensive conjugated system. The lowest intercepts belong to the cations containing only sp³-hybridized carbons and nitrogens, *i.e.* 1-alkyl-1-methylpyrrolidinium [C$_k$mpyrr]⁺, 1-alkyl-3-methylpiperidinium [C$_k$mpip]⁺ and 1-alkyl-1-methylmorpholinium [C$_k$mmor]⁺. In order to study the influence of charge delocalization in the intercepts, we have quantified the extent of charge delocalization in each heterocycle by means of the MCBO index [87], whose value for each cation is shown in Table 4. As expected, aliphatic cations have lower MCBO than aromatic cations. In fact, the aliphatic pyrrolidinium, piperidinium, or morpholinium cations present similar MCBOs despite their different structures. The same holds for the aromatic cations imidazolium, pyridinium, thiazolium and quinolinium. Counter-intuitively, the biggest aromatic cation [C$_k$quin]⁺ has not the largest MCBO despite showing the largest MPD$_i$. Nevertheless, the cations with high MCBO have higher intercepts in Fig. 2a than the low MCBO cations.

The molecular polarizabilities of the cations with an ethyl side chain ($k$ = 2) are shown as a function of the wavelength $\lambda = 2\pi c/\omega$ in Fig. 2b. Interestingly, the aromatic cations show a stronger correlation between the wavelength and the polarizability than the aliphatic cations as visible in the curvature at low wavelengths. From the Kramers-Kronig relationship [52], this increase of electronic polarizability indicates the existence of an UV resonance, which, according to previous publications [92], should shift towards the visible range as charge delocalization increases in the cations. Regarding the polarizability ranking, it differs from the MCBO (or MPD$_i$) rankings as the overall molecular volume plays a decisive role.

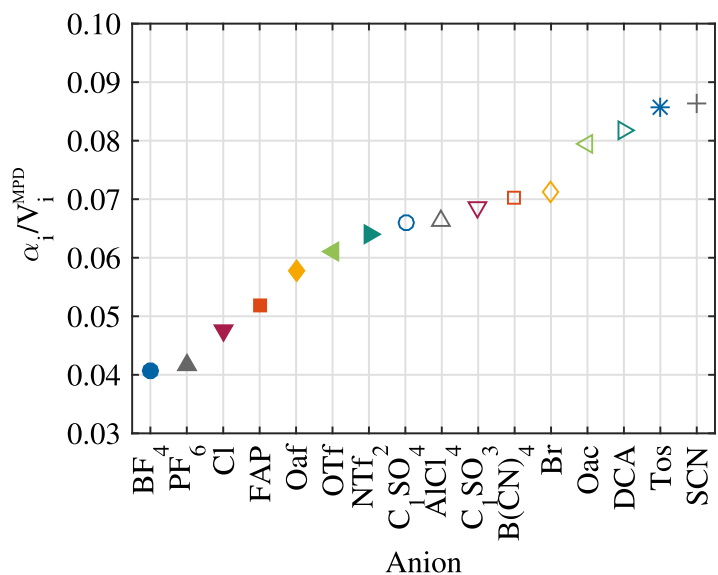

**Fig. 1.** Molecular polarizability density MPD$_i$ of the most common anions in ionic liquids.

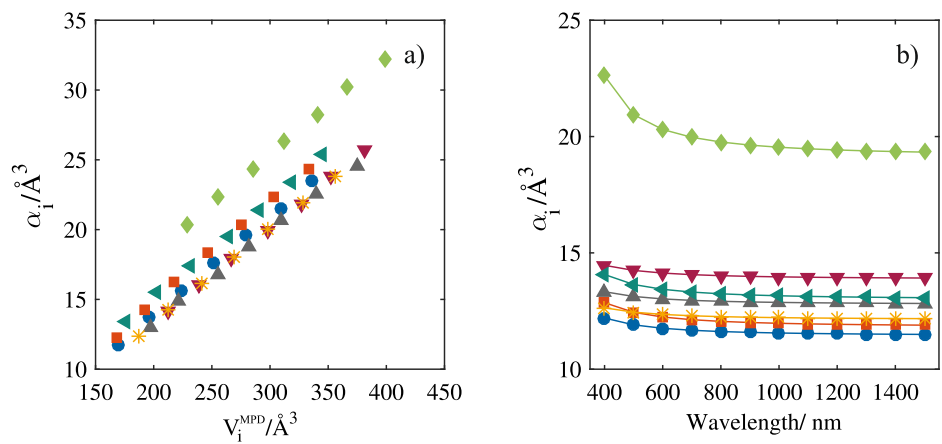

**Fig. 2.** a) Electronic polarizability versus molecular volume for different families of alkyl-heterocyclic cations. Members of the same family only differ in the length $k$ of their alkyl chain. Marker legend: $[C_k\text{mim}]^+$ (●), $[C_k\text{mmor}]^+$ (▲), $[C_k\text{mpip}]^+$ (▼), $[C_k\text{py}]^+$ (■), $[C_k\text{mpyrr}]^+$ (∗), $[C_k\text{quin}]^+$ (◆) and $[C_k\text{mthia}]^+$ (◀). b) Electronic polarizability dispersion, $\alpha(\lambda)$, for the cations with alkyl chain $k = 2$.

**Table 4**
Normalized Multi-Center Bond Order (MCBO) indices throughout the heterocycles of the considered cations bearing the same alkyl chain length ($k = 2$) and corresponding MPD$_i$. For $[C_2\text{quin}]^+$ cation the MCBO was calculated on the perimeter involving both cycles. The atomic polarizabilities $\alpha_{ji}$ are taken from [91].

| Cation | \|MCBO\| | MPD$_i$ | $\alpha_{ji}$ (N) [Å$^3$] | $\alpha_{ji}$ (C$_{sp}^2$) [Å$^3$] | $\alpha_{ji}$ (C$_{sp}^3$) [Å$^3$] |
|---|---|---|---|---|---|
| $[C_2\text{mpyrr}]^+$ | 0.28 | 0.0714 | | | |
| $[C_2\text{mpip}]^+$ | 0.30 | 0.0723 | 1.36 | | 0.84–1.11 |
| $[C_2\text{mmor}]^+$ | 0.31 | 0.0718 | | | |
| $[C_2\text{mim}]^+$ | 0.58 | 0.0745 | 1.15 | 1.03–1.12 | 0.89–1.05 |
| $[C_2\text{mthia}]^+$ | 0.59 | 0.0822 | | | |
| $[C_2\text{quin}]^+$ | 0.59 | 0.0970 | | | |
| $[C_2\text{py}]^+$ | 0.63 | 0.0783 | 1.33 | 1.13–1.22 | 0.98–1.15 |

Further evidence of the influence of charge delocalization in the electronic polarizability of heterocycles can be indirectly found in other published works. For instance, in Ref. [63] a linear decomposition of the electronic polarizability of ILs in atomic contributions distinguishing between carbon hybridization types yield higher electronic polarizability for the C(sp$^2$) species, mostly appearing in regions of charge delocalization of IL-forming ions, than for C (sp$^3$), the representative hybridization of carbon in saturated alkyl chains. Another example can be found in [82], where decomposition of the electronic polarizability in additive atomic contributions was carried out for different ionic species. In this case, it was found that the N and C(sp$^2$) atoms in the pyridinium cation have a more considerable contribution to the electronic polarizability than those of the imidazolium cation [82,91]. This difference in contributions could be produced by the linearization of the total electronic polarizability of both rings, which, according to our results, is strongly influenced by their different extent of aromaticity. However, the atomic polarizability of the nitrogen in $[C_2\text{mpip}]^+$ was found to be higher than that of the nitrogen atoms in the ring of the imidazolium cation (see Table 4 and [91]) although the former cation presents an aliphatic ring system. A difference between the sp$^2$ and sp$^3$ carbons in the imidazolium and pyridinium exists, but it is not significant.

### 3.3. Influence of the side chains on the polarizability

In this section, we study the effect that different substituents have on electronic polarizability. For this, we consider the cations of group 2 (see Table S2 in the Supporting Material). They consist of an imidazolium cation singly and doubly substituted by different groups containing $\pi$ orbitals, some of them presenting a certain degree of charge delocalization.

The first thing to highlight is that doubling the substituents leads to an increase of the MPD$_i$ compared to the singly substituted imidazolium cation. The various side chains can be grouped in three categories as shown in Table 5: The first category contains aliphatic, ethoxy, and ethylnitrile chains. The first two lack of $\pi$ orbitals in the side chain, and the third contains sp carbon instead of sp$^3$ carbons. Cations with these chains have an MCBO of roughly 0.33, and the MPD$_i$ is around 0.7. The second category gathers cations with side chains containing different number of $\pi$ orbitals on sp$^2$ carbons. The MCBO is approximately 0.35, and the MPD$_i$ increases with increasing size of the $\pi$ systems, *i.e.* the benzylic system has the largest polarizability per volume. A clear example of the impact that a simple isolated double bond has on electronic polarizability is observed in the butyl and crotyl groups. They differ only in the presence of a double bond, but it is enough to increase the crotyl group's electronic polarizability over its saturated equivalent. Hence, the contribution to the electronic polarizability of the extra $\pi$ orbital clearly overcompensates the loss of the $\sigma$ orbitals of the two hydrogen atoms that takes place upon unsaturation.

**Table 5**
Normalized Multi-Center Bond Order (MCBO) and MPD$_i$ of imidazoliums having different side chains.

| Cation | \|MCBO\| | MPD$_i$ |
|---|---|---|
| $[C_4\text{mim}]^+$ | 0.33 | 0.0700 |
| $[\text{eomim}]^+$ | 0.33 | 0.0695 |
| $[1,3\text{-dieoim}]^+$ | 0.34 | 0.0706 |
| $[\text{ennim}]^+$ | 0.34 | 0.0693 |
| $[1,3\text{-dienim}]^+$ | 0.32 | 0.0702 |
| $[\text{allylmim}]^+$ | 0.35 | 0.0733 |
| $[1,3\text{-diallylim}]^+$ | 0.37 | 0.0745 |
| $[\text{crotylmim}]^+$ | 0.32 | 0.0754 |
| $[\text{benzylmim}]^+$ | 0.36 | 0.0791 |
| $[1,3\text{-dibenzylim}]^+$ | 0.35 | 0.0820 |
| $[\text{vinylmim}]^+$ | 0.40 | 0.0757 |
| $[1,2\text{-divinylmim}]^+$ | 0.39 | 0.0826 |
| $[1,3\text{-divinylim}]^+$ | 0.48 | 0.0839 |

The third category comprises vinylic side chains. Here, the MCBO is significantly increased to 0.4 and higher. Also, the $MPD_i$ values are very high. This may indicate that the charge delocalization strengthens the polarizability as in these cations, the $\pi$ system of the ring is connected to the $\pi$ system of the side chain. Doubling the vinyl substituent yields the cation with the largest $MPD_i$, the [1,3-divinylim]$^+$. The position of this doubling induces differences in the $MPD_i$ of the cation, as shown by the different value of [1,2-divinylim]$^+$, suggesting a certain influence of the substitution on the delocalization throughout the imidazolium ring.

In order to discuss the different impact of $\sigma$ and $\pi$ orbitals have on the $MPD_i$, we turn our attention to the third group of cations. It comprises four different families of cations bearing functionalized side chains of variable length. These side chains are: one perfluorinated chain, where the unit is $-CF_2-$, one oxygenated chain, where the unit is an oxyethyl group, $-CH_2-O-CH_2-$, one polyenylic chain, where the basic unit is a carbon that is respectively linked to its neighbors by a double and a single bond, $-CH=$ or $=CH-$, and one alkyl chain where the basic unit is $-CH_2-$. Note that for all the non-oxygenated chains, $k = 0$ refers to 1-H-3-methylimidazolium cation, while for the oxygenated chain, it refers to 1-methyl-3-methylimidazolium (see Supplementary Material). In Fig. 3a the polarizability dispersion for the different chains with $k = 4$ is shown. The order of units yielding from lowest to highest polarizabilities is: $-CH_2- < -CF_2- < -CH= < -CH_2-O-CH_2-$. The highest value is yielded by the oxygenated chain, which makes sense because its basic unit contains more atoms than the other basic units. On the other hand, Fig. 3b shows the $MPD_i$ of the different cations as a function of the chain length. The order from lowest to highest $MPD_i$ is: $-CF_2- < -CH_2- \approx -CH_2-O-CH_2- < -CH=$. The figure shows that both perfluorinated and oxygenated chains decrease the overall $MPD_i$ of the imidazolium cation as their length increases, although the perfluorinated chain seems to be more efficient for this purpose. The reason for this decrease is that the ratios between electronic polarizability and volume of the O and F atoms are respectively lower than those of C and H.

It is imperative to highlight the intense influence of charge delocalization in the $MPD_i$ of the polyenylic chain. Except for

$k = 1$, it shows much higher values than those of the other chains, which are composed of larger units and exclusively present $\sigma$ bonds. The significant increase of the $MPD_i$ with $k$ in Fig. 3b suggests the breakdown of the hypothesis of linear polarizability. In order to analyze this behaviour in a more detailed way, the electronic polarizabilities and molar volumes of this family of cations, [uC$_k$mim]$^+$, is compared with its saturated equivalent, [C$_k$mim]$^+$, in Fig. 4. The [C$_k$mim]$^+$ family contains a standard alkyl chain where bonds between carbons present a clear $\sigma$ character, which is associated with a robust charge localization. Nevertheless, the polyenyl chain of [uC$_k$mim]$^+$ presents $\pi$ conjugation, and thus, electronic delocalization throughout all the chain length. As it is clearly visible in Fig. 4a, molecular volume $V_i^{MPD}$ is a linear function of the chain length $k$ in both families of cations. However, different slopes are registered in each case because double bonds are roughly 0.2 Å shorter than single bonds. Furthermore, the saturated side chain contains one more hydrogen per carbon. In the case of electronic polarizabilities, as shown in Fig. 4b, the [C$_k$mim]$^+$ family presents a linear trend with the chain length while the [uC$_k$mim]$^+$ family not. In [C$_k$mim]$^+$, each $-CH_2-$ unit seems to yield a uniform contribution to the overall polarizability, which is the expected behavior since each $-CH_2-$ unit can be identified as an electronically independent region. Specifically, each $-CH_2-$ unit is made of $\sigma$ orbitals with a highly localized electronic charge and, in consequence, according to Eq. (2), the overall polarizability of the chain results in the sum of $k$ independent contributions of size $L$. This fact was exploited in predicting molecular polarizabilities in our former studies [62,63].

Nevertheless, the polyenyl chain does not show this linear behavior anymore. In terms of Eq. (2), each new unit in the polyenylic chain increases the extension of the delocalization region emerging from the $\pi$ conjugated system. Hence, the effective $L$ in the equation is a linear function of the chain length $k : L_{effective} \approx k \cdot L$. Henceforth, according to Eq. (2), a non-linear increase of electronic polarizability should be observed, and that is, in fact, what the simulations of the polyenyl chain show. As a result, the prediction based on uniform atomic contributions as postulated in [62,63] breaks down. This can also be deduced from the atomic polarizabilities in Fig. 5.

The atomic polarizabilities are decomposed into polarization ($\alpha_\beta^p$) and charge transfer ($\alpha_\beta^{CT}$) contributions. The local character of the polarizability is determined by the polarization contribution $\alpha_\beta^p$. It shows only a slight increase with increasing chain length. Interestingly, there is a zigzag behavior of the carbons C7 to C13 in the aliphatic side chain. The steep increase in the molecular polarizability $\alpha_i$ emerges from the charge transfer contribution $\alpha_\beta^{CT}$ of the side-chain carbons. For example, the atomic polarizabil-ity of the carbon C7 increases from 1.00 Å$^3$ for $k = 2$ to 2.35 Å$^3$ for $k = 8$ rendering any prediction algorithm based on Designed Regression [62,63] obsolete. Interestingly, the atomic polarizability $\alpha_i^{CT}$ for each sp$^2$ carbon in the side chain seems to follow a parabolic curve. The atomic polarizability in the center of the conjugated system is a maximum and decreases towards its ends as visible in Fig. 5a. The conjugated system extends to the heterocyclic core of the imidazolium as the nitrogen N3 and the carbon C4 of the imidazolium ring increase their atomic polarizability with increasing side chain length $k$. However, the effect is significantly smaller than the observed on the side-chain carbons C7-C14.

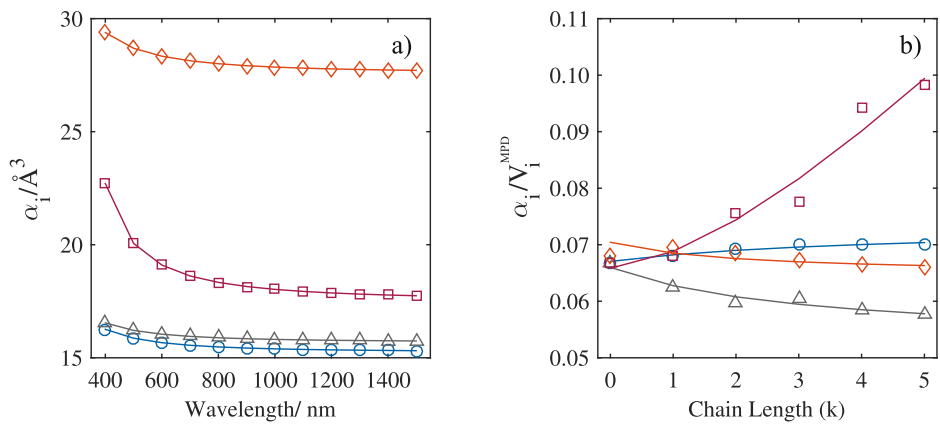

**Fig. 3.** a) Electronic polarizability dispersion, $\alpha(\lambda)$, for the imidazolium cation bearing the functionalized chains of length $k = 4$. b) $MPD_i$ at $\lambda = 589$ nm as a function of the chain length, $k$. Marker legend: $-CH_2-$ ($\circ$), $-CF_2-$ ($\triangle$), $-CH=$ (or $=CH-$) ($\square$) and $-CH_2-O-CH_2-$ ($\diamond$).

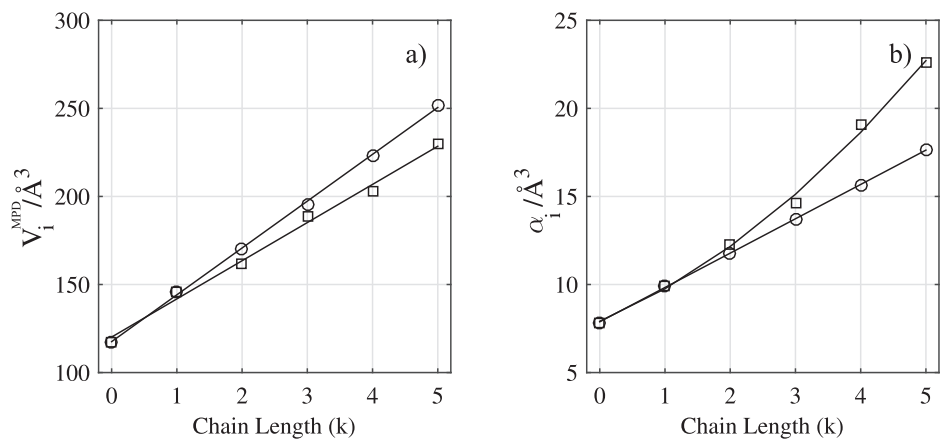

**Fig. 4.** a) Molecular volume and b) electronic polarizability as a function of the number of carbons in (∘) the alkyl chain of a [C$_k$mim]$^+$ cation and in (□) the polyenyl chain of [uC$_k$mim]$^+$ cation. The key difference among both chains is the existence of conjugation along the polyenyl substituent in contrast to the saturated character of the alkyl one.

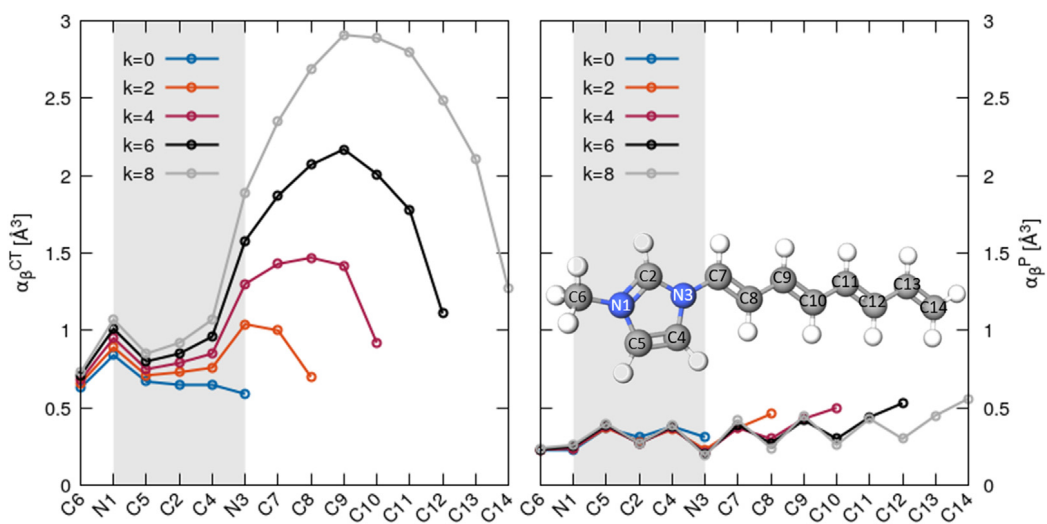

**Fig. 5.** Decomposition of the molecular polarizability of [uC$_k$mim]$^+$ into atomic contributions. The lines have no physical meaning and are a guide for the eye only.

The overall polarizability of the [uC$_k$mim]$^+$ family as a function of the chain length $k$ in Fig. 4b can be extrapolated by

$$\alpha_i(k) = c_i \cdot \left(V_i^{MPD}(k)\right)^{\gamma_i} \tag{14}$$

$$MPD_i(k) = c_i \cdot \left(V_i^{MPD}(k)\right)^{\gamma_i - 1} \tag{15}$$

Here, we also used the fact that the molecular volume $V_i^{MPD}(k)$, see Fig. 4a, is a linear function of the chain length $k$. The constant $c_i$ depends on the nature of the molecule. Nevertheless, with increasing chain length $k$, the $\gamma$-effect of the side chain becomes more important than that of the core unit. The corresponding parameters can be found in Table 6. Charge delocalization effects are present if $\gamma_i$ are above one. This threshold is already visible in Fig. 3b. [F$_k$-mim]$^+$ and [m(eo)$_k$mim]$^+$ have a $\gamma_i < 1$ and their MPD$_i$ decrease with increasing chain length $k$. However, [C$_k$mim]$^+$ and [uC$_k$mim]$^+$ possess $\gamma_i > 1$, and, therefore, their MPD$_i$ increase with the chain length. Nevertheless, a significant effect is only visible for [uC$_k$-mim]$^+$, whereas the effect for [C$_k$mim]$^+$ seems to level-off.

### 3.4. Designing the refractive index based on cation/anion combinations

Based on our results, we are able to compute the refractive index according to the Lorentz-Lorenz Eq. (1). We obtained and combined the molecular polarizabilities $\alpha_i$ and volumes $V_i^{MPD}$ of the most common cations and anions to produce a map of the estimated refractive index of 1216 ion pairs, Fig. 6. In the map, ions are ordered according to their MPD$_i$ values (increasing from top to bot-tom left to right). The quantitative values of the refractive index of each ion combination can be found in Table S4 of the Supplementary Material. The accuracy of our estimations were compared with available experimental data. 30% of the ILs were predicted with an absolute error below 0.005 in the refractive index and about 50% with an error below 0.01 (see Fig. S1 in the Supplementary Material). The refractive index of more than 75% of the ILs composing the test group was predicted with an absolute error below 0.02, and 87% was predicted with an error below 0.03. In addition, there is a set of ILs whose refractive indices are not predicted by our model. This set mainly comprises ILs including mono-atomic halogen ions like [Cl]⁻ and [Br]⁻. Halogen-based ILs comprise about 9.9% of the entire IL test set. The wrong evaluation of their refrac-tive indices could be attributed to strong electronic polarizability deviations due to bulk effects or due to our volume treatment, which could fail when considering mono-atomic ions. Thus, discarding this set of ions, the error we commit when predicting the refractive index is close to the expected one according to the calibration method described in the Theoretical and Computational Details section. In accordance, this map is expected to be a powerful tool to introduce ILs in the design of photonic devices. Finally, some trends can be discussed. The sequence on the x-axis on the map corresponds to the $MPD_i$ ranking in Fig. 1. As expected and visible by the color code in Fig. 6, the refractive index $n$ for a given cation follows most often the corresponding $MPD_i$ of the anions.

**Table 6**
Delocalization parameters of several side chains.

| Cation | $c_i$ | $\gamma_i$ |
|---|---|---|
| [F$_k$mim]$^+$ | 0.121 | 0.870 |
| [m(eo)$_k$mim]$^+$ | 0.088 | 0.954 |
| [C$_k$mim]$^+$ | 0.051 | 1.057 |
| [uC$_k$mim]$^+$ | 0.002 | 1.730 |

However, there are exceptions to this finding such as those of ILs based on the [FAP]⁻ anion. These ILs present refractive indices markedly lower than the ones which would correspond them according to their position in the $MPD_i$ ranking. This is because the volume $V_{IL}$ is not the sum of the molecular volumes $V_i^{MPD}$, but a volume cross-term has to be taken into account, Eq. 9. As [FAP]⁻ is the biggest anion under investigation, its cross-term term is the largest, and consequently, it has a larger impact on the resulting refractive indices.

Many high refractive index compounds with $n > 2.0$ are solid at room temperature, *e.g.* AsI₃ or SnI₄. In 2006, Seddon et al. tried to make ionic liquids with such high refractive indexes [93]. However, only those based on the polyiodides [I₇]⁻ and [I₉]⁻ reached that threshold value. In a theoretical study, some of us proposed to exchange the cation in [C₂mim][Iₓ] with ethylammonium to decrease the polyiodides to [I₃]⁻, [24]. Notwithstanding, the map in Fig. 6 opens a new path for achieving high refractive index ILs by exploiting extensive charge delocalization on the cations. According to Eq. (14), the refractive index threshold of 2.0 can also be achieved by long conjugated chain imidazoliums with [SCN]⁻ or [Tos]⁻ as visible in Fig. 7. The ionic liquid combinations proposed here have the advantage that the anions are stable. Polyiodides tend to decompose into [I₂]⁻ and smaller iodide anions. For the [uCₖmim]⁺ cations, the molar volume is a linear function with the chain length, but the electronic polarizability is not. In consequence, the refractive index increases with increasing chain length $k$. However, it was experimentally found that conjugated systems also reach a saturation regime when the number of repetition units is large enough [72], fact which is not taken into account in Fig. 7. Interestingly, very high refractive indices are even possible with hydrophobic [FAP]⁻ anions using the cation's charge delocalization effect. On the other hand, in imidazoliums with saturated side chains, the number of —CH₂— units should be as low as possible to have significant refractive indices.

## 4. Conclusions

In this paper, we report the effect that the presence of delocalized electrons in conjugated systems on the molecular structures of ionic liquid-forming ions induces in their electronic polarizability. The usual assumption of uniform, additive contributions of atoms or molecular groups in the molecules is seen to be invalid for this particular case, because of the practical impossibility to define virtually independent intramolecular groups. Moreover, we report a more accurate treatment of quantum-mechanical volumes derived from the electron density evaluation, leading to significantly improved predictions of volume-based properties of large ionic liquid ions. We applied these more accurate volumes to the prediction of the refractive index, but also other volume-based properties such as density can also be predicted with increased precision. Moreover, we report a modified fractional relation between polarizability and volume.

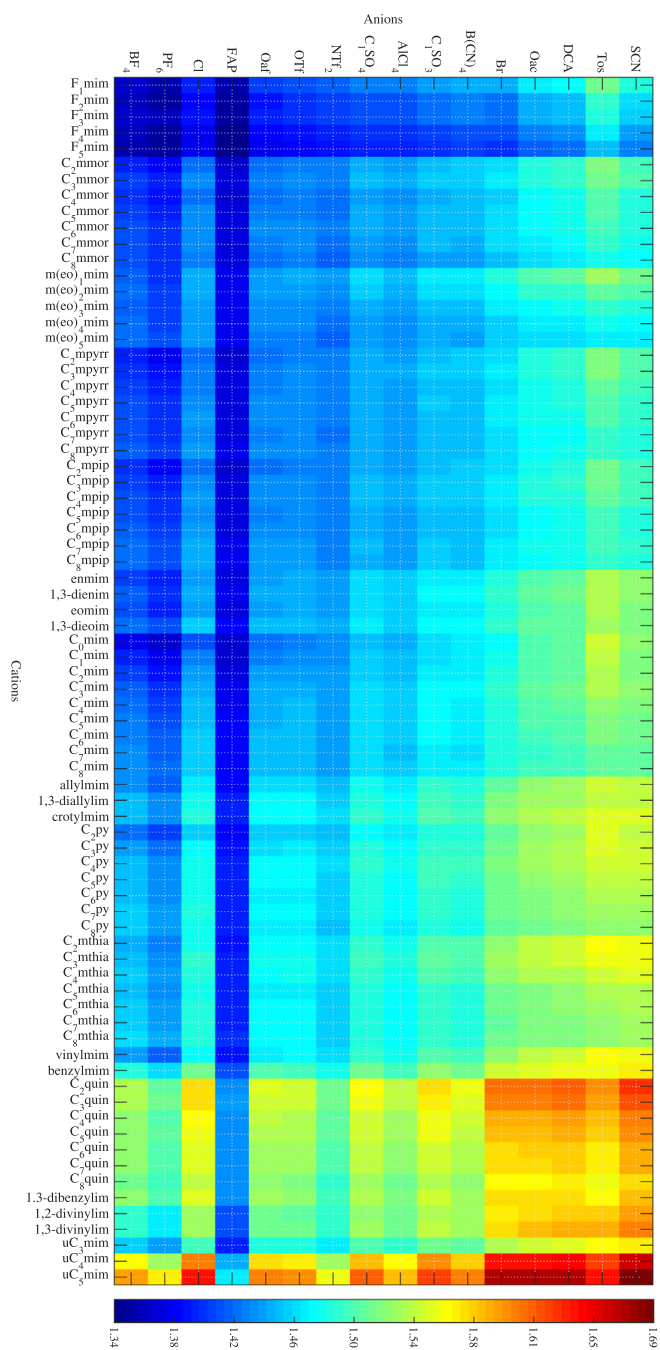

**Fig. 6.** Map of the predicted refractive index of the combinations of the different ions at $\lambda = 589$ nm. Meaning of abbreviations can be found in Tables of the supplementary material.

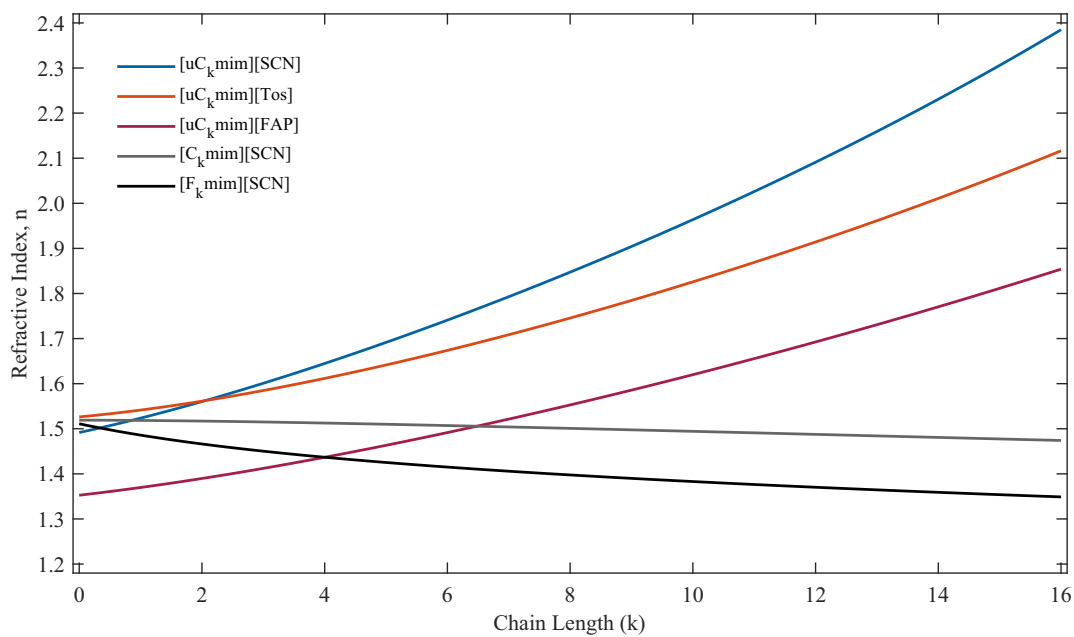

**Fig. 7.** Prediction of the refractive index as a function of the cationic chain length *k*.

On the other hand, in the past, the most considerable effort in making ionic liquids with high refractive indexes has evolved around the design of optimal anions with highly polarizable atoms. However, in the present work, we showed that increasing the fraction of delocalized electrons in the ionic liquid molecular species (e.g. including imidazolium cations with long, conjugated side chains) high refractive indexes can be achieved. Indeed, the carbon polarizability in conjugated carbon side chains is not uniform as expected by former theories but exhibits a parabolic behavior. Here, carbons in the middle of the side chain may have atomic polarizabilities that are more than twice the corresponding terminal carbons' value, and the charge delocalization effect also penetrates the aromatic ring. Finally, a map of the refractive index of the 1216 combinations of the studied anions and cations is reported, providing the geography of high polarizability regions of the ionic liquid world.

### CRediT authorship contribution statement

**Carlos Damián Rodríguez-Fernández:** Conceptualization, Methodology, Software, Validation, Formal analysis, Investigation, Data curation, Visualization, Funding acquisition, Writing - original draft, Writing - review & editing. **Elena López Lago:** Conceptualization, Methodology, Validation, Formal analysis, Investigation, Supervision, Funding acquisition, Writing - original draft, Writing - review & editing. **Christian Schröder:** Conceptualization, Methodology, Software, Validation, Formal analysis, Investigation, Visualization, Writing - original draft, Writing - review & editing. **Luis M. Varela:** Conceptualization, Methodology, Validation, Formal analysis, Investigation, Supervision, Funding acquisition Project administration, Writing - original draft, Writing - review & editing.

### Declaration of Competing Interest

The authors declare that they have no known competing financial interests or personal relationships that could have appeared to influence the work reported in this paper.

### Acknowledgements

This work was supported by Ministerio de Economia y Competitividad (MINECO) and FEDER Program through the project MAT2017-89239-C2-1-P; Xunta de Galicia and FEDER (ED431D 2017/06, ED431E2018/08, GRC 508 ED431C 2020/10). C. D. R. F. thanks the support of Xunta de Galicia through the grant ED481A-2018/032. We also thank the Centro de Supercomputacion de Galicia (CESGA) facility, Santiago de Compostela, Galicia, Spain, for providing the computational resources employed in this work.

### Appendix A. Supplementary material

Supplementary data associated with this article can be found, in the online version, at https://doi.org/10.1016/j.molliq.2021. 117099.